\documentclass[conference]{IEEEtran}

\usepackage{xcolor} 
\usepackage{url}
\usepackage{graphicx}
\usepackage{svg}
\ifCLASSINFOpdf
\else
\fi

\usepackage{xspace}

 \newcommand{\acname}[1]{\ensuremath{\mathsf{#1}}\xspace}
 \newcommand{\layname}[1]{\ensuremath{\mathsf{#1}}\xspace}
 \newcommand{\pktname}[1]{\texttt{#1}}

\usepackage{graphicx}

\begin{document}
%
\title{Breaking 5G on The Lower Layer}

\author{\IEEEauthorblockN{Subangkar Karmaker Shanto}
	\IEEEauthorblockA{Purdue University\\
		sshanto@purdue.edu}
	\and
	\IEEEauthorblockN{Imtiaz Karim}
	\IEEEauthorblockA{The University of Texas at Dallas\\
		imtiaz.karim@utdallas.edu}
	\and
	\IEEEauthorblockN{Elisa Bertino}
	\IEEEauthorblockA{Purdue University\\
		bertino@purdue.edu}}

\IEEEoverridecommandlockouts
\makeatletter\def\@IEEEpubidpullup{6.5\baselineskip}\makeatother
\IEEEpubid{\parbox{\columnwidth}{
		Workshop on Security and Privacy of Next-Generation Networks \\ (FutureG) 2026 \\
		23 February 2026, San Diego, CA, USA \\
		ISBN 979-8-9919276-9-7\\ 
		https://dx.doi.org/10.14722/futureg.2026.240039 \\   
		www.ndss-symposium.org
}
\hspace{\columnsep}\makebox[\columnwidth]{}}

\newcommand{\sks}[1]{\textcolor{purple}{\small\textbf{[Subangkar]}#1$\triangleleft$}}

\maketitle

\begin{abstract}
As 3GPP systems have strengthened security at the upper layers of the cellular stack, plaintext {PHY} and {MAC} layers have remained relatively understudied, though interest in them is growing.
In this work, we explore lower-layer exploitation in modern 5G, where recent releases have increased the number of lower-layer control messages and procedures, creating new opportunities for practical attacks. 
We present two practical attacks and evaluate them in a controlled lab testbed. First, we reproduce a {SIB1} spoofing attack to study manipulations of unprotected broadcast fields. By repeatedly changing a key parameter, the UE is forced to refresh and reacquire system information, keeping the radio interface active longer than necessary and increasing battery consumption. Second, we demonstrate a new Timing Advance (TA) manipulation attack during the random access procedure. By injecting an attacker-chosen TA offset in the random access response, the victim applies incorrect uplink timing, which leads to uplink desynchronization, radio link failures, and repeated reconnection loops that effectively cause denial of service.
Our experiments use commercial smartphones and open-source 5G network software. Experimental results in our testbed demonstrate that TA offsets exceeding a small tolerance reliably trigger radio link failures in our testbed and can keep devices stuck in repeated re-establishment attempts as long as the rogue base station remains present. Overall, our findings highlight that compact lower-layer control messages can have a significant impact on availability and power, and they motivate placing defenses for initial access and broadcast procedures.

\end{abstract}


%
\IEEEpeerreviewmaketitle

\section{Introduction}
Security analysis of upper layers of the cellular stack of 4G and 5G have been a field of research for a long time. A lot of research works focus on the upper layers of the cellular stack, including \layname{RRC} and \layname{NAS}~\cite{tu2024logic, touchinguntouchables, LTEphonecatcher,practicalattacks,breakinglayertwo, hussain2019reasoner, hussain2025controlchaos, lteinspector}. Accordingly, the defense has been studied thoroughly for those layers. Consequently, 3GPP has strengthened the security of
5G increasing subscriber-identity privacy via SUPI concealment (SUCI) and applies NAS security after the security mode procedure~\cite{ts33501}. However, the introduction of encryption mechanisms at control layers comes at a cost of higher latency~\cite{michaelides2025assessing}. Hence, to facilitate faster reconfiguration of the network many \layname{PHY}/\layname{MAC} procedures in current 4G/5G systems are neither encrypted nor integrity protected~\cite{3gpp.38.331, ludant2025wiselowlayer}. This creates scope for adversaries to exploit the users. 
However, prior works have predominantly focused on vulnerabilities in 4G/5G upper-layer protocols and services, such as core network functions, \layname{RRC}/\layname{NAS} signaling, and network slicing, while comparatively fewer studies examine low-layer (\layname{PHY}/\layname{MAC}) control procedures~\cite{zhang2022airinterface, ahmad2022systematic5g}. A recent work has explored this direction of exploiting the lower layers of 4G/5G~\cite{ludant2025wiselowlayer} motivating further exploration of this direction.

In this work, we present a systematic evaluation for unprotected control-plane parameters in 5G NR, identifying critical vulnerable fields across \layname{MAC} and \layname{RRC} layers that lack integrity protection during initial access and broadcast procedures. Building upon prior work~\cite{ludant2025wiselowlayer} that speculated potential vulnerabilities in Timing Advance Command manipulation and performed \pktname{SIB1} spoofing,  we implement and empirically validate these attacks through rogue gNodeB deployment combining passive reconnaissance via downlink sniffing and active manipulation using software-defined radio platforms. 
Our systematic experimental evaluation demonstrates that Timing Advance Command manipulation in Random Access Response messages causes connected-mode denial-of-service through uplink timing desynchronization and Radio Link Failure, while \pktname{SIB1} spoofing induces idle-mode battery drain through forced system information re-acquisition cycles. Additionally, we systematically explore Tracking Area Code modification and SI window length toggling attacks to assess their practical feasibility.
\section{Related Works}
Though security analyses of 3GPP networks have largely centered on \layname{RRC} and \layname{NAS} layers where cryptographic protections exist, recent works have emphasized that the \layname{PHY} and \layname{MAC} layers remain \emph{unprotected}, enabling energy-efficient spoofing, resource abuse, and passive privacy leakage. \emph{Ludant et al.}~\cite{ludant2025wiselowlayer} demonstrate systematic exploitation of beamforming, DCI, and \layname{MAC} CE messages for sub-20 m localization, induced jamming, HARQ desynchronization, and carrier aggregation misuse in both 4G and 5G, motivating further lower layer security analysis.

Multiple recent studies characterize and exploit similar low-layer weaknesses. \emph{Shaik et al.} and subsequent \layname{PHY}/\layname{MAC} injection works on 4G LTE highlight the practicality of control channel spoofing and adaptive overshadowing to degrade link availability~\cite{practicalattacks, adaptOver}. In 5G, NR-specific telemetry and decoder tools further show the feasibility of capturing temporary identifiers and unprotected scheduling procedures, expanding the attack surface for \layname{MAC}-layer abuse~\cite{nrscope}.

Recent 4G/5G research also targets physical and \layname{MAC}-layer vulnerabilities in real deployments. \emph{Cao et al.} specifically examine 5G \layname{MAC} Control Elements, revealing that integrity-less scheduling and mobility signaling can be tampered to force misallocation or sustain energy-draining states~\cite{cao2025mac}. Complementary surveys highlight jamming, spoofing, beam manipulation, and ML-based detection challenges at the \layname{PHY}/\layname{MAC} layers in 5G, LTE and wireless networks~\cite{vu2024phyply, hossein2022jamming, zhang2025pls}.
Finally, while prior work demonstrated generic \layname{PHY}/\layname{MAC} spoofing including resource abuse and beam leakage, the \emph{TA \layname{MAC} CE injection} threat was only discussed as a potential vector for uplink timing desynchronization. In addition, there is no work that experimentally evaluated additional unprotected broadcast or scheduling fields. Our work moves beyond speculation by providing the first empirical evidence that \emph{over the air TA manipulation in 5G NR} can break \layname{PHY} timing alignment, trigger RLF, and sustain \layname{RLC}/\layname{RRC} reconnection loops, exposing a concrete and previously unvalidated low layer vulnerability~\cite{ludant2025wiselowlayer}. Building on their insights, we also systematically analyze additional integrity-less broadcast and scheduling fields through \pktname{SIB1} spoofing to expand the known 4G/5G unprotected control-plane attack surface.

\section{Background}
\subsection{5G NR}
5G New Radio (NR) organizes the radio interface into three main layers: L1 (\layname{PHY}), L2 (which includes \layname{MAC}, \layname{RLC} and \layname{PDCP}), and L3 (\layname{RRC}) \cite[Sec.~4.4]{TS38300}. At a high level, \layname{PHY} (L1) handles the actual radio waveform (for example, synchronization and transmission/reception on physical channels), while \layname{MAC} (part of L2) coordinates shared access to the air interface by scheduling and multiplexing control/data over time--frequency resources~\cite[Sec.~4.4]{ts38321}. \layname{RRC} (L3) is the main access-stratum control protocol between the UE and gNB; it configures radio resources and bearers, supports mobility and measurements, and manages key connection-level control functions~\cite[Sec.~4.4]{3gpp.38.331}. Above these, the Non-Access Stratum (\layname{NAS}) is the UE-to-core control protocol (for example, with the AMF), and its messages are carried through the RAN using \layname{RRC} 
signaling~\cite[Sec.~7.6]{TS38300}.

A typical control-plane entry flow starts when the UE finds a cell, synchronizes, reads broadcast system information, and then runs Random Access to obtain uplink timing alignment and initial uplink resources~\cite[Sec.~5.3.4, Sec.~9.2.6]{TS38300}. In contention-based Random Access, the \layname{MAC} layer coordinates the key steps (preamble transmission, random access response reception, scheduled uplink transmission, and contention resolution) before declaring the procedure 
complete~\cite[Sec.~5.1]{ts38321}. Once access succeeds, the UE establishes an \layname{RRC} connection to enable dedicated signaling (and subsequent security activation and configuration) \cite[Sec.~5.3.3]{3gpp.38.331}. With the \layname{RRC} connection in place, the UE can execute \layname{NAS} procedures with the core network, such as 5G registration~\cite[Sec.~5.5.1]{ts24501}.

\subsection{System Information Block} Following the broader role of system information in 5G NR, we now focus on the primary broadcast block that governs UE initial access behavior. The System Information Block Type 1 (\pktname{SIB1}) serves as the fundamental broadcast information container in 5G NR networks, transmitted periodically on the Physical Downlink Shared Channel to provide essential cell access parameters to all UEs regardless of their connection state. According to 3GPP TS 38.331~\cite{3gpp.38.331}, \pktname{SIB1} carries critical configuration including PLMN identities, Tracking Area Code, cell barring information, RACH configuration, and common channel parameters required for initial cell selection and access. UEs in \emph{RRC\_IDLE} and \emph{RRC\_INACTIVE} states continuously monitor \pktname{SIB1} transmissions by decoding \acname{PDCCH} with SI-RNTI (System Information RNTI) to detect scheduling grants for \pktname{SIB1} on \acname{PDSCH}, typically transmitted with a periodicity of 160 milliseconds.
Central to the change notification mechanism, \pktname{SIB1} contains the \textit{valueTag} field (a 5-bit parameter ranging from 0 to 31) that acts as a version indicator for system information content. When the network modifies any system information parameters, it increments \textit{valueTag} (modulo 32) to signal UEs that cached system information has become stale. The detection of a changed \textit{valueTag} can be used by the UE to judge whether stored SI remains valid to ensure they operate with current network configuration. This mechanism enables efficient broadcast updates without requiring dedicated signaling to individual UEs, but critically assumes integrity of the value tag parameter to prevent unnecessary acquisition cycles.

The Tracking Area Code (TAC) field: $trackingAreaCode$, within \pktname{SIB1}'s cellAccessRelatedInfo specifies the 24-bit tracking area identifier for mobility management. UEs compare the received TAC against their registered tracking area; mismatches trigger Tracking Area Update procedures to maintain location registration with the 5G Core AMF, enabling efficient paging and mobility tracking across network coverage areas. Additionally, the \textit{si-WindowLength} parameter in \pktname{SIB1}'s si-SchedulingInfo defines the time window duration during which on-demand System Information messages (\pktname{SIB2}-\pktname{SIB9}) are transmitted on \acname{PDSCH}. UEs use this value to calculate precise \acname{PDCCH} monitoring occasions for SI-RNTI-based scheduling grants, establishing deterministic timing for acquiring cell-specific configuration beyond minimum \pktname{SIB1} content.

\subsection{Timing Advance Value}
In 5G New Radio (NR) networks, the Timing Advance (TA) mechanism maintains its critical role as a foundational synchronization primitive at the Medium Access Control (\layname{MAC}) layer, ensuring temporal alignment of uplink transmissions within the Orthogonal Frequency Division Multiple Access (OFDMA) framework. The necessity of TA in 5G NR becomes even more pronounced compared to LTE due to the introduction of flexible numerologies, higher frequency bands, and stringent latency requirements.

\begin{figure}[ht]
    \centering
    \includegraphics[width=0.7\columnwidth]{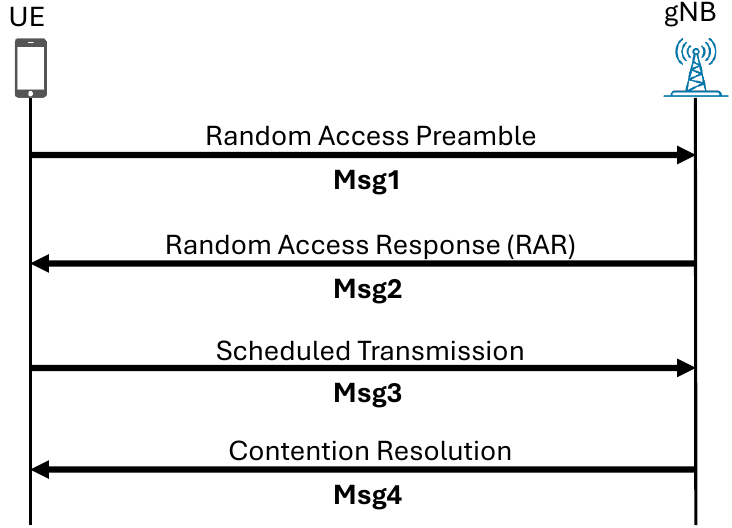}
    \caption{RACH procedure for initial cell access}
    \label{fig:rach}
\end{figure}

In the Random Access Channel (RACH) procedure specified in 3GPP TS 38.321~\cite{ts38321} as shown in Figure~\ref{fig:rach}, the UE transmits Msg1 containing a randomly selected preamble sequence from a configured set on the Physical Random Access Channel (\acname{PRACH}), similar to LTE but with enhanced preamble formats supporting different cell sizes and subcarrier spacings. The gNodeB (gNB) performs preamble detection through correlation processing, identifying both the preamble index and its timing offset relative to the expected \acname{PRACH} reception window. This timing offset directly reflects propagation delay between UE and gNB, ranging from near-zero for small cells to several hundred microseconds for macro deployments. The gNB quantizes this measured offset into a Timing Advance Command, where each TA unit in 5G NR corresponds to a numerology-dependent time quantum to maintain consistent physical timing resolution across numerologies.

The Random Access Response (\pktname{RAR}), transmitted as Msg2 in RACH, carries the computed TA command from gNB to UE via a \layname{MAC} Protocol Data Unit (PDU) on the Physical Downlink Shared Channel (\acname{PDSCH}). The \layname{MAC} \pktname{RAR} payload in 5G NR contains a 12-bit Timing Advance Command field (expanded from LTE's 11 bits) capable of representing values from 0 to 3846, providing finer granularity and extended range to accommodate diverse deployment scenarios from dense urban small cells to rural macro cells spanning tens of kilometers. Upon successful decoding of the \pktname{RAR}, identified by monitoring the Physical Downlink Control Channel (PDCCH) for Downlink Control Information (DCI) format $1\_0$ scrambled with the Random Access Radio Network Temporary Identifier (RA-RNTI) computed from the \acname{PRACH} transmission occasion parameters including time-frequency resource index, slot number, etc. The UE extracts the TA command and adjusts its uplink timing parameter NTA. This adjustment shifts the UE's entire uplink transmission timing forward by the specified amount, ensuring that subsequent uplink transmissions, beginning with Msg3 (containing \layname{RRC} Setup Request in the procedure) and continuing through all Physical Uplink Shared Channel (\acname{PUSCH}) and Physical Uplink Control Channel (\acname{PUCCH}) transmissions, arrive at the gNB aligned to slot boundaries within the tolerance provided by the cyclic prefix.

According to 3GPP TS 33.501~\cite{ts33501}, the 5G security architecture implements integrity protection and ciphering of user-plane and control-plane data only after successful completion of the primary authentication procedure in the 5G Core Network, which necessarily occurs after RACH completion since radio resource establishment and \layname{RRC} connection setup are prerequisites for authentication messaging. Consequently, the Random Access procedure encompassing Msg1 through Msg4 is executed without integrity protection, creating an exploitable attack surface at the physical and \layname{MAC} layers. This design decision, inherited from previous cellular generations and maintained due to bootstrapping constraints where UEs lack established security context during initial network access, allows adversaries to manipulate critical synchronization parameters without cryptographic verification.

\section{Systematic Analysis of Lower Layer Parameters}

\begin{table*}[t]
\centering
\caption{Selected Unprotected Control-Plane Parameters}
\label{tab:parameter_selection}
\begin{tabular}{|l|l|l|p{7cm}|}
\hline
\textbf{Parameter} & \textbf{Layer} & \textbf{Message} & \textbf{Role} \\
\hline
Timing Advance (TA) Command & \layname{MAC} & \pktname{RAR} & Controls uplink timing synchronization by specifying transmission advance to compensate propagation delay \\
\hline
\textit{valueTag} & \layname{RRC} & \pktname{SIB1} & Version indicator for system information content; increment triggers complete SI re-acquisition \\
\hline
Tracking Area Code (TAC) & \layname{RRC}/\layname{NAS} & \pktname{SIB1} cellAccessRelatedInfo & Identifies tracking area for mobility management; mismatch triggers Tracking Area Update to AMF \\
\hline
\textit{si-WindowLength} & \layname{RRC} & \pktname{SIB1} si-SchedulingInfo & Defines time window duration (5/10/15/20 ms) for scheduling on-demand SI messages (\pktname{SIB2}-\pktname{SIB9}) \\
\hline
\end{tabular}%
\end{table*}

We systematically selected parameters on \acname{PDSCH} to explore based on their lack of integrity protection, direct control over resource-intensive UE operations, and broadcast impact scope, with part of the selection rationale motivated by~\cite{ludant2025wiselowlayer} on lower-layer desynchronization and mobility signaling attacks. Table~\ref{tab:parameter_selection} lists the explored fields. We prioritize parameters that are (i) pre-integrity, (ii) critical to broadcast/initial access, and (iii) induce extra UE control procedures (e.g., SI reacquisition/mobility) or impact uplink timing. The \textit{valueTag} of \pktname{SIB1} triggers computationally expensive system information re-acquisition affecting battery life, Tracking Area Code controls mobility signaling to core network, SI window length governs scheduling calculations for on-demand broadcasts, and Timing Advance Command manages uplink synchronization at the \layname{PHY}/\layname{MAC} layer where misalignment causes radio link failures. Together, these fields provide systematic coverage of the unprotected control-plane attack surface.

\subsection{Threat Model}
We consider an adversary located within the radio coverage region of a 4G/5G base station (\textbf{BS}). The adversary is assumed to possess a Software Defined Radio (\textbf{SDR}) platform and is capable of passively capturing RF I\&Q sample streams from nearby cellular cells. Furthermore, consistent with prior work~\cite{OWL,FALCON,lteeye,5gsniffer, ludant2025wiselowlayer}, we assume that the adversary can decode unencrypted BS downlink channels, including but not limited to synchronization and broadcast signals, system information blocks, scheduling grants, and low layer control headers. We name this decoder as sniffer. Finally, building on  capabilities demonstrated in the literature~\cite{ adaptOver,hidingplainsignal,LTrack,ludant2025wiselowlayer}, the adversary is assumed to be able to synthesize and inject valid 4G/5G waveforms over the air with fine-grained temporal control, enabling targeted signal transmission at attacker specified time instants. Our testbed evaluation uses a rogue gNB/fake base station (FBS) to realize the required over-the-air downlink injection capability in a controlled setting. We do not evaluate ``overshadowing'' like~\cite{hidingplainsignal} against a live serving gNB in our experiment.


\subsection{Root Causes of Vulnerability.}
We leverage the absence of integrity protection as per 3GPP TS 33.501~\cite{ts33501} in 5G broadcast and initial access procedures. The UE continuously monitors \pktname{SIB1} on \acname{PDSCH}, as required by 3GPP compliance, even if the actual configuration is unchanged. Since \pktname{SIB1} must be decoded before security context establishment, it lacks cryptographic integrity protection, enabling a rogue base station to modify parameters undetected and trigger false change notifications. The same vulnerability applies to Timing Advance (TA) extraction in the \pktname{RAR}, which occurs prior to \layname{RRC} security activation; the UE applies the malicious TA to its physical layer uplink timing reference and sends Msg3 on the granted \acname{PUSCH} resource using the injected offset, shifting all subsequent uplink transmissions, including the \layname{RRC} Setup Request. Persistent uplink desynchronization can trigger RLF and reconnection loops.

\subsection{SIB1 Spoofing Attack}
\noindent \textbf{Attack Overview.} We systematically evaluate manipulation of three unprotected \pktname{SIB1} broadcast parameters—\textit{valueTag}, \textit{trackingAreaCode}, and \textit{si-WindowLength}. Our experiments use a rogue gNodeB (fake base station) to perform controlled over-the-air injection and to assess denial-of-service and resource-exhaustion effects on UEs in \emph{RRC\_IDLE} and \emph{RRC\_INACTIVE} states that monitor system information. UEs in \emph{RRC\_CONNECTED} typically do not reacquire \pktname{SIB1} unless explicitly triggered by the network (e.g., paging), so we focus on idle/inactive states.
\begin{figure}[ht]
    \centering
    \includegraphics[width=0.98\columnwidth]{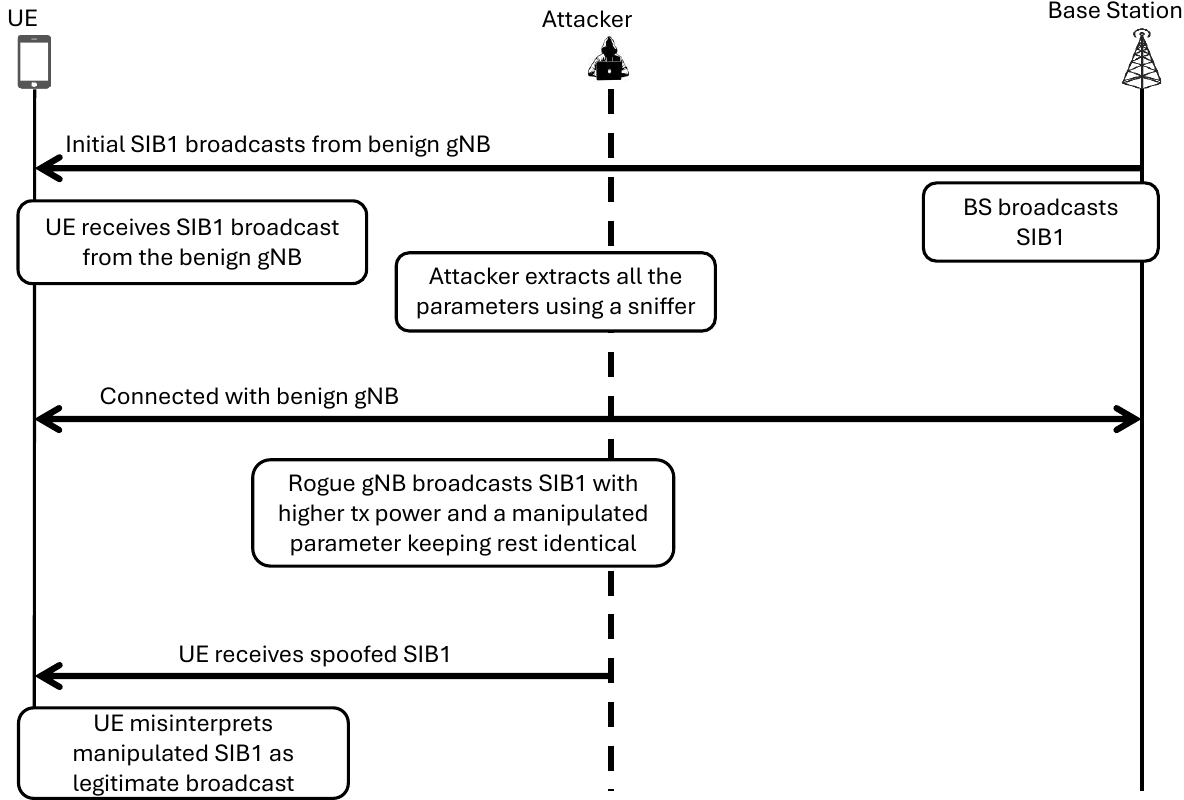}
    \caption{\pktname{SIB1} Spoofing}
    \label{fig:sib1_attack}
\end{figure}

\noindent \textbf{Attack Steps.} We next describe the procedure for modifying each selected field in the \pktname{SIB1} message for analysis using the \pktname{SIB1} spoofing attack from~\cite{ludant2025wiselowlayer} as shown in Figure~\ref{fig:sib1_attack}.

\noindent \textit{\textit{valueTag} Manipulation:} After capturing legitimate \pktname{SIB1} parameters via downlink sniffing from Primary and Secondary Synchronization Signals and Physical Broadcast Channel, the rogue gNB periodically increments the \textit{valueTag} field (5-bit value ranging 0-31) every 10 seconds while maintaining all other \pktname{SIB1} content unchanged. Upon detecting the modified value tag through periodic SI-RNTI monitoring on Physical Downlink Control Channel, the victim UE interprets this as system information change notification and discard cached SI. The UE then initiates complete system information acquisition by continuously monitoring \acname{PDCCH} Control Resource Set zero for Downlink Control Information Format 1\_0 with CRC scrambled by SI-RNTI, performing blind decoding to locate scheduling grants. Once detected, the UE decodes the scheduled Physical Downlink Shared Channel transmission containing \pktname{SIB1}, applies LDPC decoding for transport block extraction, only to discover no meaningful parameter changes beyond the incremented tag forcing repeated unnecessary acquisition cycles.

\noindent \textit{Tracking Area Code Modification:} The rogue gNB periodically modifies the 24-bit Tracking Area Code field within \pktname{SIB1}'s cellAccessRelatedInfo information element every 30 seconds after initial Parameter harvesting via \acname{PDSCH} decoding of legitimate broadcasts, cycling through different TAC values to simulate UE movement across tracking area boundaries. According to 3GPP TS 24.501, a UE upon detection of TAC mismatches between cached registration area and received \pktname{SIB1} while camping in \emph{RRC\_IDLE} mode may trigger a registration update when the UE determines it is camping on a TA not in its stored TAI list involving \layname{NAS} Registration Request messages encapsulated in \layname{RRC} UplinkInformationTransfer and transmitted via Physical Uplink Shared Channel to initiate N1 signaling with the Access and Mobility Management Function in 5G Core Network, potentially generating cascading signaling storms as multiple victim UEs simultaneously respond to broadcast TAC modifications with individual mobility management transactions consuming core network processing resources.

\noindent \textit{SI Window Length Toggling:} The rogue gNB alternates the \textit{si-WindowLength} parameter in \pktname{SIB1}'s si-SchedulingInfo information element between enumerated values ms5, ms10, and ms20 across periodic \pktname{SIB1} transmissions on Physical Downlink Shared Channel with 160 ms periodicity, disrupting the deterministic scheduling framework defined in TS 38.331 Section 5.2.2.3 for on-demand System Information Blocks. UEs calculate exact Physical Downlink Control Channel monitoring occasions for additional SI messages (\pktname{SIB2}-\pktname{SIB9}) based on the standard formula. When the window length toggles between monitoring cycles for example from ms10 to ms20, the UEs applying previously cached values miscalculate SI-RNTI-scrambled \acname{PDCCH} monitoring occasions within Control Resource Sets, causing blind decoding attempts in incorrect time-frequency resources while missing actual SI transmission slots, forcing extended \acname{PDCCH} monitoring across multiple SI windows and potential system information acquisition failures for cell-edge UEs with marginal Signal-to-Interference-plus-Noise Ratio.

\noindent \textbf{Expected Impact.}
Frequent modification of \textit{valueTag} may damage UE's battery life because it systematically disrupts the Discontinuous Reception power-saving mechanism that enables energy-efficient idle-mode operation. Under normal conditions, idle-mode UEs enter deep sleep states between paging occasions, waking briefly to monitor paging channels and periodically verify system information validity, achieving duty cycles below 1 percent with receiver circuitry powered down more than 99 percent of the time. The continuous \textit{valueTag} modifications force UEs to remain in active receiver mode for extended periods performing computationally intensive operations. Alongside individual UE battery drain, the attack may generate sustained Physical Downlink Control Channel congestion as all victim UEs simultaneously attempt SI acquisition following each value tag modification, competing for limited \acname{PDCCH} blind decoding capacity and potentially blocking detection of legitimate paging messages or other control signaling.
Tracking Area Code modification aims to trigger mobility management signaling storms by forcing victim UEs to initiate Tracking Area Update procedures upon detecting TAC mismatches, causing each UE to transmit \layname{NAS} Registration Request messages via Physical Uplink Shared Channel to the Access and Mobility Management Function. 
When multiple victim UEs simultaneously respond to broadcast TAC changes, this might generate localized signaling load on the serving AMF, though large-scale overload would require coordinated attacks across multiple base stations given typical AMF dimensioning for large subscriber populations. In an overshadowing setting against a legitimate network, broadcast TAC changes can increase NAS update attempts toward the core; in our rogue gNB (standalone) evaluation, the effect is primarily UE-side activity toward the rogue cell.
SI window length toggling may disrupt deterministic scheduling for on-demand system information acquisition by causing UEs to miscalculate Physical Downlink Control Channel monitoring occasions when cached \textit{si-WindowLength} values mismatch current broadcasts. This might force extended \acname{PDCCH} monitoring across multiple windows, delay acquisition of critical system information (\pktname{SIB2} cell reselection parameters, \pktname{SIB4} neighboring cell lists), and cause service disruption for cell-edge UEs with marginal SINR where scheduling misalignment compounds into complete SI acquisition failure.

\subsection{TA Manipulation Attack}
\noindent \textbf{Attack Overview.} In this attack we manipulate the Timing Advance (TA) value during the RACH procedure by an attacker-chosen offset with the intention to cause denial of service on the victim UE. 
Compared to a fake base station (FBS) or rogue gNB that simply attracts UEs and provides no service, TA manipulation breaks availability by desynchronizing uplink timing via the (integrity-less) \pktname{RAR} TA field. In contrast to inactive rogue gNBs, TA manipulation maintains a partially functional connection: the UE completes RACH, believes it is connected, and repeatedly attempts recovery. This can cause failures after access appears to progress, leading to repeated RLF/retry loops, rather than an immediate ``no service'' failure.
\begin{figure*}[ht]
    \centering
    \includegraphics[width=0.79\textwidth]{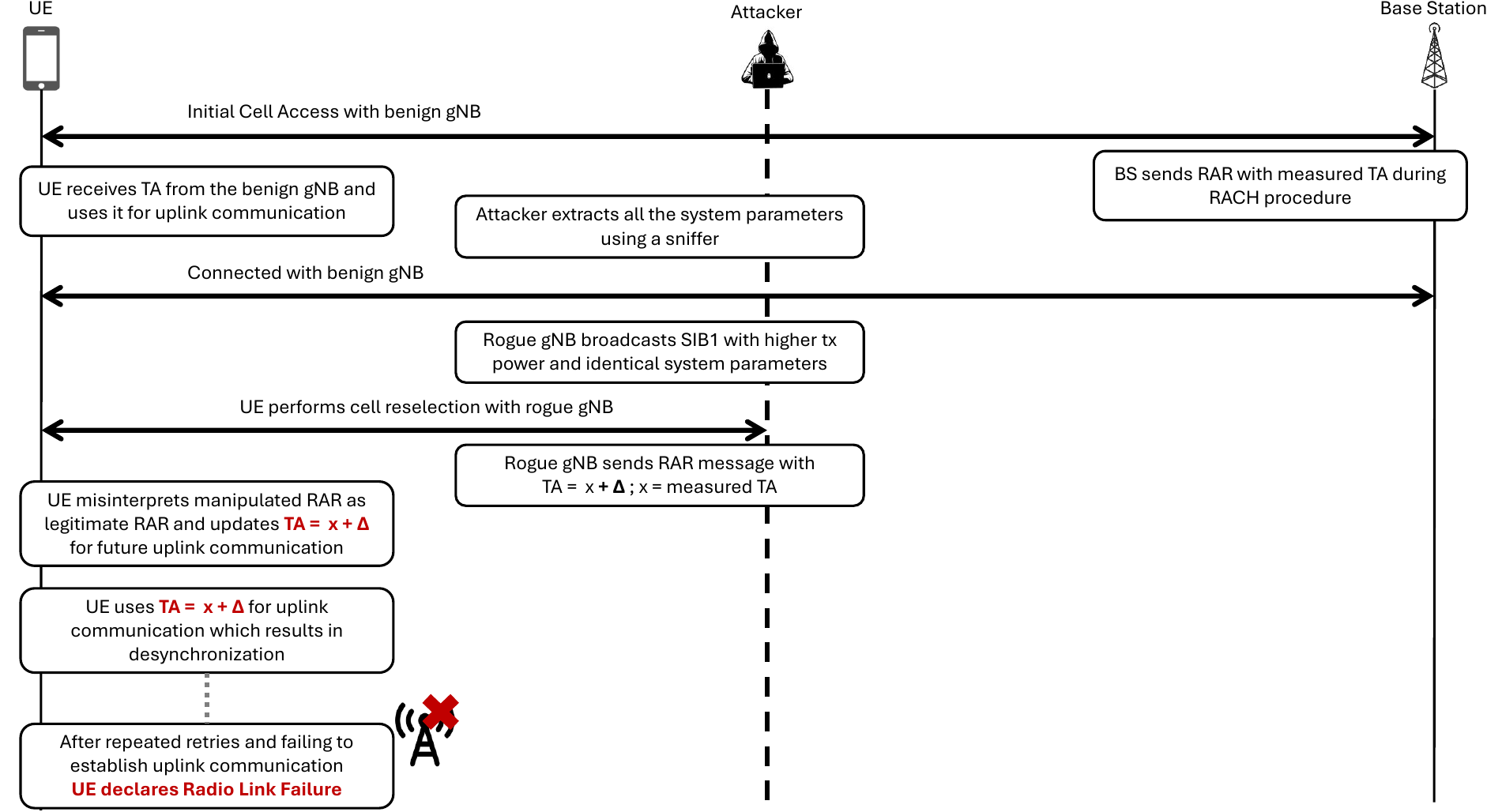}
    \caption{TA Manipulation causes uplink de-synchronization and leads to Radio Link Failure (RLF)}
    \label{fig:ta_attack}
\end{figure*}

\noindent \textbf{Attack Steps.} Figure~\ref{fig:ta_attack} demonstrates our attack. Our attack methodology operates in two phases: passive reconnaissance through downlink sniffing to harvest system parameters followed by active manipulation via rogue gNB deployment replicating legitimate cell configurations to attract victim UEs. 
In the reconnaissance phase, the sniffer synchronizes to the target cell by detecting and decoding the Primary Synchronization Signal (PSS) and Secondary Synchronization Signal (SSS) transmitted in every Synchronization Signal Block (SSB), which occurs periodically with configurable periodicity. From the detected Physical Cell Identity (PCI) encoded in PSS/SSS and the Master Information Block (MIB) carried on the Physical Broadcast Channel (\acname{PBCH}) within each SSB, the sniffer extracts critical system parameters including System Frame Number (SFN), subcarrier spacing for common control channels, \acname{PDCCH} configuration via ControlResourceSet (CORESET) \#0, and frequency location of initial downlink bandwidth part. 
With successful SSB detection and MIB decoding, the sniffer proceeds to decode System Information Block 1 (\pktname{SIB1}) scheduled on \acname{PDSCH} according to timing and frequency allocation rules defined in 3GPP TS 38.213~\cite{3gpp.38.213}. \pktname{SIB1} contains essential cell configuration parameters including Public Land Mobile Network (PLMN) identities, cell identity, Tracking Area Code, cell barring status, RACH configuration parameters, common channel configurations, and serving cell configuration for initial access. Of particular importance for the attack, \pktname{SIB1}'s RACH-ConfigCommon information element specifies critical Random Access procedure parameters: the set of available \acname{PRACH} preamble formats supporting different cell sizes and subcarrier spacings, \acname{PRACH} frequency and time resource allocation defining which resource blocks and slots carry \acname{PRACH} occasions, preamble transmission power parameters specifying target received power at gNB, power ramping parameters controlling power increase across consecutive RACH attempts, and RA response window configuration defining the duration during which UEs monitor \acname{PDCCH} for \pktname{RAR} following preamble transmission.
Additionally, the sniffer monitors \acname{PDCCH} transmissions to observe Downlink Control Information scheduling \pktname{RAR} messages. By decoding DCI format transmissions with CRC scrambled by RA-RNTI values corresponding to observed \acname{PRACH} occasions, the sniffer can capture subsequent \acname{PDSCH} transmissions carrying \pktname{RAR} \layname{MAC} PDUs. Analysis of these legitimate \pktname{RAR} messages reveals typical Timing Advance Command values employed by the target gNB for UEs at various distances.

Armed with comprehensive system information and operational parameters extracted during reconnaissance, the adversary transitions to the active attack phase by deploying a rogue gNB. The adversary configures the rogue gNB with identical PLMN identities extracted from legitimate \pktname{SIB1} during reconnaissance, ensuring victims perceive it as belonging to their home network or authorized roaming partners, avoiding automatic rejection due to PLMN mismatch. Crucially, the rogue gNB replicates all essential system parameters and transmits with higher signal power with a shorter distance from the UE than the benign gNB. This replication ensures that victim UEs, upon selecting the rogue cell based on favorable RSRP measurements, find expected system information enabling successful initial access procedures without triggering protocol inconsistencies.

When a victim UE selects the rogue gNB and initiates the Random Access procedure by transmitting a \acname{PRACH} preamble on Msg1, the rogue base station performs standard preamble detection to identify the transmitted preamble index and measure its timing offset. This measured timing offset represents the legitimate propagation delay requiring compensation through timing advance. At this point, the attack diverges from legitimate protocol operation through injection of a manipulated timing offset into the Random Access Response. The rogue gNB computes the manipulated timing advance command by adding an offset to the legitimate measured value. The magnitude and sign of this offset determine attack severity and impact characteristics: positive values cause premature UE transmission with uplink signals arriving earlier than intended subframe boundaries, while negative values induce delayed transmission with late-arriving signals. The adversary selects offset values based on attack objectives, balancing several considerations. 
Finally, the manipulated timing advance value is encoded into the Random Access Response \layname{MAC} PDU structure defined in 3GPP TS 38.321~\cite{ts38321}, which contains \layname{MAC} subheader identifying the \pktname{RAR} for the detected preamble index and \layname{MAC} \pktname{RAR} payload comprising the Timing Advance Command field, frequency domain resource assignment for Msg3 uplink grant, time domain resource assignment, Modulation and Coding Scheme for Msg3, and Temporary C-RNTI assignment. The rogue gNB assembles a properly formatted \pktname{RAR} with the manipulated timing advance value while maintaining correct structure for all other fields, ensuring the victim UE successfully decodes and processes the message. The \pktname{RAR} is transmitted on \acname{PDSCH} in the time window specified by the RACH response window parameter, scheduled by DCI format on \acname{PDCCH} with CRC scrambled by the appropriate RA-RNTI value computed from the \acname{PRACH} occasion parameters matching the victim's Msg1 transmission.

From the victim UE's perspective, the received \pktname{RAR} appears entirely legitimate and indistinguishable from authentic network responses. The \pktname{RAR} arrives within the expected response window following \acname{PRACH} transmission, carries appropriate RA-RNTI scrambling matching the transmitted preamble occasion, and contains correctly formatted \layname{MAC} subheader and payload fields passing all protocol validity checks.\\
\noindent \textbf{Expected Impact.}
The physical layer consequences of timing advance manipulation in 5G NR might be substantially severe. A manipulated TA value may disrupt uplink transmission alignment, potentially driving the UE into a persistent loss of uplink synchronization. When the induced timing offset exceeds the gNB's tolerance for symbol boundary alignment, repeated decoding failures on scheduled uplink channels (e.g., \acname{PUSCH} or \acname{PUCCH}) prevent successful reception of control feedback and data. As a result, the gNB ceases uplink scheduling, marking the UE as out of sync. The UE, observing consecutive HARQ feedback failures and the absence of valid uplink grants, subsequently declare a Radio Link Failure (RLF). This transitions the UE to \emph{RRC\_IDLE} and initiates cell reselection followed by a new RACH procedure. Under sustained TA manipulation, the UE can reenter \emph{RRC\_CONNECTED} only to encounter recurring uplink timing misalignment, forming a reconnection loop characterized by repeated RLF declarations and RACH retries. Such loops amplify signaling overhead, increase random access congestion, and degrade service continuity, effectively creating a \textbf{denial of service} (DoS) condition through forced desynchronization and repeated connection reestablishment.

\section{Experiment and Evaluation}
\subsection{Evaluation Setup}
\subsubsection*{Hardware}
Our experimental testbed comprises software-defined radio (SDR) devices utilizing USRP B210 hardware platforms for both legitimate and rogue gNB implementations. Each USRP B210 connects via USB 3.0 to dedicated laptop systems with low-latency kernel configurations, equipped with Intel Core Ultra 7 processors and 32GB RAM. For UE, we used commercial off-the-shelf (COTS) smartphones including OnePlus Nord 5G (Qualcomm Snapdragon 765G with X52 5G modem supporting FR1 bands n1/n3/n7/n28/n78/n78) and Google Pixel 5 (Snapdragon 765G with identical modem capabilities), both running Android Operating System without custom radio interface modifications, representing realistic victim devices operating according to 3GPP standard UE behavior. The testbed operates in controlled indoor laboratory environment with RF shielding to prevent interference with commercial networks.

\subsubsection*{Software}
The legitimate base station deployment consists of integrated BS and Core network software components executing on a single host machine running Ubuntu 22.04 LTS, implementing complete end-to-end network functionality including Access and Mobility Management Function, Session Management Function, and User Plane Function elements. In contrast, the rogue base station operates in standalone mode executing only Radio Access Network protocol stack without core network connectivity, running on identical hardware specifications. Consequently, any NAS/control signaling impact on a real core network is not exercised in our evaluation. The base station software implementations utilize the open-source srsRAN~\cite{srsRAN} codebase for physical layer and protocol stack functionality, with the legitimate deployment using unmodified release version while the rogue implementation incorporates custom modifications to \layname{MAC} layer Random Access Response generation and System Information Block broadcasting procedures as detailed in previous sections. The testbed evaluation was conducted using one concurrently active UE. The setup models simultaneous uplink timing behavior and generalizes to multi-UE scenarios without altering the validity of the experiment conditions. 
Real-time protocol monitoring and packet inspection at the UE side utilizes Network Signal Guru (NSG) application~\cite[ver.~4.7.10]{networksignalguru} with root privileges enabled through Magisk framework, providing access to Qualcomm diagnostic interface for capturing physical layer parameters, \layname{RRC} messages, \layname{MAC} PDUs, and timing advance values during Random Access procedures and connected mode operation.

\subsection{Experimental Results}
\subsubsection{SIB1 Spoofing Attack Result}
Our experimental validation of the \pktname{SIB1} spoofing attack demonstrates increased system-information reacquisition activity across both test UE models under sustained \textit{valueTag} manipulation. With the rogue base station configured to increment \textit{valueTag} every 10 seconds while maintaining constant transmission power 5 dB higher than the legitimate network to ensure victim cell selection, both OnePlus Nord 5G and Google Pixel 5 devices exhibited increased SI reacquisition activity, which is expected to reduce standby time compared to baseline. 
Network Signal Guru logs confirmed the attack mechanism, showing \pktname{SIB1} acquisition events occurring at precise 10-second intervals with \textit{valueTag} values incrementing sequentially, while actual \pktname{SIB1} content parameters including PLMN-IdentityInfo, cellAccessRelatedInfo, and connEstFailureControl remained constant across all acquisitions. This confirms the successful demonstration of the attack in our testbed.
In contrast, Tracking Area Code modification and SI window length toggling attacks demonstrate limited practical impact in our testbed. Despite successful TAC modification in broadcast \pktname{SIB1} and confirmed UE detection through Network Signal Guru logging showing updated TAC values in decoded cellAccessRelatedInfo fields, commercial UEs (OnePlus Nord 5G and Pixel 5) did not trigger expected Tracking Area Update procedures or generate \layname{NAS} signaling to the Core. Implementation-specific optimizations may defer mobility updates until user-plane traffic demand. Similarly, SI window length toggling between ms5, ms10, and ms20 values produced no observable system information acquisition failures or extended monitoring delays in our testbed. These negative results suggest that while these broadcast parameters lack integrity protection and remain theoretically manipulable, their exploitation requires either specific network topology conditions, edge-case propagation scenarios (low SINR for SI window), or alternative attack methodologies beyond simple parameter toggling to manifest observable impact.

\subsubsection{TA Manipulation Attack Result}

Our experimental evaluation of the Timing Advance (TA) manipulation attack demonstrates severe disruption to UE connectivity under controlled manipulation parameters. Upon activating the rogue BS with manipulated Random Access Response transmission, both the OnePlus Nord 5G and the Google Pixel 5 devices exhibited Radio Link Failure within 30 to 60 seconds of initial RACH procedure completion, with timing variations attributable to UE-specific RLF detection thresholds and timer configurations per 3GPP TS 38.331~\cite{3gpp.38.331}. We systematically evaluated $TA\_delta$ injection values ranging from 5 to 60 timing advance units to characterize the empirically observed TA-delta tolerance range (for the tested UEs). At $TA\_delta \leq 10$, both tested UE models maintained stable connectivity without observable service degradation, indicating that this manipulation magnitude remains within the combined tolerance of cyclic prefix protection and UE receiver robustness margins. However, increasing $TA\_delta$ to values exceeding 20 units consistently triggered Radio Link Failure for both device models, with $TA\_delta$ values of 30-60 units producing reliable RLF within the observed window. The RLF triggering mechanism follows the standard 3GPP procedure where persistent uplink timing misalignment causes repeated \acname{PUSCH} and \acname{PUCCH} decoding failures at the BS, resulting in accumulated out-of-sync indications at the UE physical layer, subsequently starting timer T310, and declaring Radio Link Failure upon T310 expiration without synchronization recovery.

Following RLF declaration, both UE models initiated \layname{RRC} re-establishment procedures as specified in TS 38.331~\cite[Sec.~5.3.7]{3gpp.38.331}, attempting to recover the radio connection by performing cell selection and transmitting \emph{RRCReestablishmentRequest} messages. However, due to the rogue BS maintaining higher transmission power for Synchronization Signal Blocks compared to the legitimate network, the UEs consistently reselected the rogue cell during the re-establishment procedure rather than reverting to the legitimate BS as shown in Figure~\ref{fig:ta_attack}. Consequently, subsequent RACH attempts during re-establishment again received manipulated Random Access Response messages with identical $TA\_delta$ injection, perpetuating the timing misalignment condition. This failure loop persisted for the entire duration that the rogue gNB was active in our testbed. We did not observe an automatic “escape” (e.g., cell blacklisting) within our observation window (more than an hour) for both the UEs we experimented with; such vendor-specific mitigation, if present, may bound persistence and is left to future work. Network Signal Guru packet captures confirmed this cyclic behavior, showing repeated \emph{RRCReestablishmentRequest} transmissions at intervals corresponding to RLF detection periods, with each re-establishment attempt receiving manipulated Timing Advance Commands visible in \layname{MAC} \pktname{RAR} subheader decoding. The UEs remained trapped in this connection failure loop for the entire duration of rogue BS operation, effectively denying stable network connectivity despite successful completion of individual RACH procedures, demonstrating the persistent denial-of-service capability of the attack as long as the rogue base station maintains transmission.\\
\subsection{Limitations}
Our evaluation uses two commercial UEs that share the same baseband family; thus, the observed TA-offset tolerance ranges and time-to-RLF should be interpreted as implementation-dependent and may vary across other modem vendors and firmware versions. For \pktname{SIB1} spoofing, we observed increased system-information reacquisition activity compared to baseline; however, we do not report power/current measurements and leave quantifying the resulting battery-life reduction (hours or \% under controlled power methodology) to future work.


\section{Conclusion}
Our TA manipulation attack empirically demonstrates that small control-plane tweaks at the \layname{MAC}/\layname{PHY} boundary can have an outsized impact on uplink alignment, pushing the UE into desynchronization and recovery loops, while our replication of the System Information Block spoofing attack validates that broadcast parameter manipulation creates increased control plane activity through forced system information re-acquisition cycles. These results highlight that lower-layer control messages, although compact and frequent, are security-critical because they directly influence radio timing, access stability, and power management. 
The absence of integrity protection for both Random Access Response messages at the \layname{MAC} layer and broadcast System Information Blocks at the \layname{RRC} layer creates exploitable vulnerabilities affecting idle-mode UEs, demonstrating that security gaps exist at the lower layers.

Practical defense against these attacks requires combining cryptographic protection with intelligent monitoring at both UE and network sides. 
Integrity protection for Random Access Response messages and broadcast \pktname{SIB1}, using keys derived from UE credentials and access parameters, can allow UEs to reject manipulated TA commands and spoofed system information before application.
Complementary UE and network-side monitoring—such as detecting inconsistent TA-signal strength relationships, abnormal \textit{valueTag} update patterns, anomalous TA distributions, or synchronized \pktname{SIB1} re-acquisition patterns, and RF fingerprinting that analyzes hardware-specific transmission characteristics—can reduce attack success rates while preserving normal 5G operation. Future work will quantify detection/false-positive rates across vendor-diverse UEs and live networks, and prototype lightweight authenticated \pktname{RAR}/\pktname{SIB1} signaling with measured overhead.

\section*{Ethics Considerations}
All experiments conducted in this study adhered to applicable ethical guidelines. The evaluations were performed in an controlled lab environment using only our own attacker and victim devices. We further ensured physical and RF isolation of the testbed, with no adjacent third-party systems present or impacted during attack execution.

\section*{Acknowledgment}
The work reported in this paper has
been supported by NSF under grant 2112471, the University of Texas System Rising STARs Award (No. 40071109), and the startup funding from the University of Texas at Dallas.




\begin{thebibliography}{1}

\bibitem{ts33501}
``{5G; Security architecture and procedures for 5G System (3GPP TS 33.501 version 16.9.0 Release 16)},'' 2022.


\bibitem{ts38321}
``{5G; NR; Medium Access Control (MAC) protocol specification (3GPP TS 38.321 version 17.3.0 Release 17)},'' 2023.


\bibitem{TS38300}
3GPP TS 38.300, ``NR; NR and NG-RAN Overall description; Stage-2,'' Rel-18, v18.4.0, Dec.~2024.

\bibitem{ts24501}
3GPP TS 24.501, ``Non-Access-Stratum (NAS) protocol for 5G System (5GS); Stage 3,'' Rel-18.

\bibitem{3gpp.38.213}
3GPP, ``NR; Physical layer procedures for control,''
3rd Generation Partnership Project (3GPP), TS~38.213, v18.5.0 (Release 18), Jan.~2025.

\bibitem{3gpp.38.331}
3GPP, ``NR; Radio Resource Control (RRC); Protocol specification,'' 
3rd Generation Partnership Project (3GPP), TS~38.331, v17.2.0, Sep.~2023.




\bibitem{tu2024logic}
K.~Tu, A.~Al Ishtiaq, S.~M.~M.~Rashid, Y.~Dong, W.~Wang, T.~Wu, and S.~R.~Hussain,
``Logic Gone Astray: A Security Analysis Framework for the Control Plane Protocols of 5G Basebands,''
in \emph{Proc. USENIX Security Symp.}, 2024.


\bibitem{touchinguntouchables}
H.~Kim, J.~Lee, E.~Lee, and Y.~Kim, ``Touching the untouchables: Dynamic security analysis of the {LTE} control plane,'' in \emph{2019 {IEEE} Symposium on Security and Privacy (SP)}, 2019.

\bibitem{LTEphonecatcher}
C.~Yu, S.~Chen, Z.~Cai, and J.~D\'{\i}az-Verdejo, ``{LTE Phone Number Catcher: A Practical Attack against Mobile Privacy},'' \emph{Security and Comm. Networks}, 2019.

\bibitem{practicalattacks}
A.~Shaik, R.~Borgaonkar, N.~Asokan, V.~Niemi, and J.-P. Seifert, ``{Practical Attacks Against Privacy and Availability in 4G/LTE Mobile Communication Systems},'' 2016.

\bibitem{breakinglayertwo}
D.~Rupprecht, K.~Kohls, T.~Holz, and C.~P\"{o}pper, ``Breaking {LTE} on layer two,'' in \emph{IEEE Symposium on Security \& Privacy (SP)}, 2019.

\bibitem{OWL}
N.~Bui and J.~Widmer, ``{OWL: A Reliable Online Watcher for LTE Control Channel Measurements},'' in \emph{Proceedings of the 5th Workshop on All Things Cellular: Operations, Applications and Challenges}, 2016.

\bibitem{FALCON}
R.~Falkenberg and C.~Wietfeld, ``{FALCON: An Accurate Real-Time Monitor for Client-Based Mobile Network Data Analytics},'' in \emph{2019 IEEE Global Communications Conference (GLOBECOM)}, 2019.

\bibitem{lteeye}
S.~Kumar, E.~Hamed, D.~Katabi, and L.~Erran, ``{LTE Radio Analytics Made Easy and Accessible},'' in \emph{Proceedings of the ACM Conference on SIGCOMM}, 2014.

\bibitem{5gsniffer}
N.~Ludant, P.~Robyns, and G.~Noubir, ``{From 5G Sniffing to Harvesting Leakages of Privacy-Preserving Messengers},'' in \emph{2023 IEEE Symposium on Security and Privacy (SP)}, 2023.

\bibitem{adaptOver}
S.~Erni, M.~Kotuliak, P.~Leu, M.~Roeschlin, and S.~Capkun, ``Adaptover: Adaptive overshadowing attacks in cellular networks,'' in \emph{Proceedings of the 28th Annual International Conference on Mobile Computing And Networking}, 2022.

\bibitem{LTrack}
M.~Kotuliak, S.~Erni, P.~Leu, M.~Roeschlin, and S.~Capkun, ``{LTrack}: Stealthy tracking of mobile phones in {LTE},'' in \emph{31st USENIX Security Symposium}, 2022.

\bibitem{hidingplainsignal}
H.~Yang, S.~Bae, M.~Son, H.~Kim, S.~M. Kim, and Y.~Kim, ``Hiding in plain signal: Physical signal overshadowing attack on {LTE},'' in \emph{28th {USENIX} Security Symposium ({USENIX} Security 19)}, 2019.

\bibitem{yongdaesniffer}
T.~D. Hoang, C.~Park, M.~Son, T.~Oh, S.~Bae, J.~Ahn, B.~Oh, and Y.~Kim, ``{LTESniffer: An Open-Source LTE Downlink/Uplink Eavesdropper},'' in \emph{Proceedings of the 16th ACM Conference on Security and Privacy in Wireless and Mobile Networks}, 2023.

\bibitem{sigunder}
N.~Ludant and G.~Noubir, ``{SigUnder: A Stealthy 5G Low Power Attack and Defenses},'' in \emph{Proceedings of the 14th ACM Conference on Security and Privacy in Wireless and Mobile Networks}, 2021.

\bibitem{ltejammingspoofing}
M.~Lichtman, R.~P. Jover, M.~Labib, R.~Rao, V.~Marojevic, and J.~H. Reed, ``{LTE}/{LTE}-{A} jamming, spoofing, and sniffing: threat assessment and mitigation,'' \emph{IEEE Communications Magazine}, 2016.

\bibitem{CIoTSignalling}
Z.~Tan, B.~Ding, J.~Zhao, Y.~Guo, and S.~Lu, ``{Data-Plane Signaling in Cellular IoT: Attacks and Defense},'' in \emph{Proceedings of the 27th Annual International Conference on Mobile Computing and Networking}, 2021.

\bibitem{CAcurrentusage}
R.~Sanchez-Mejias, Y.~Guo, M.~Lauridsen, P.~Mogensen, and L.~A. Maestro Ruiz~de Temino, ``Current consumption measurements with a carrier aggregation smartphone,'' in \emph{2014 IEEE 80th Vehicular Technology Conference}, 2014.

\bibitem{toaSRS}
A.~Blanco, N.~Ludant, P.~J. Mateo, Z.~Shi, Y.~Wang, and J.~Widmer, ``{Performance Evaluation of Single Base Station ToA-AoA Localization in an LTE Testbed},'' in \emph{2019 IEEE 30th Annual International Symposium on Personal, Indoor and Mobile Radio Communications}, 2019.

\bibitem{bsrinjection2007}
D.~Forsberg, H.~Leping, K.~Tsuyoshi, and S.~Alanara, ``{Enhancing Security and Privacy in 3GPP E-UTRAN Radio Interface},'' in \emph{IEEE 18th International Symposium on Personal, Indoor and Mobile Radio Communications}, 2007.

\bibitem{ltelocationtracking}
R.~P. Jover, ``{LTE security, protocol exploits and location tracking experimentation with low-cost software radio},'' 2016.

\bibitem{qxdm}
Qualcomm, ``{QxDM Professional Qualcomm eXtensible Diagnostic Monitor.}'' 2022.

\bibitem{srsRAN}
{SRS}, ``{Software Radio Systems. Open source SDR 4G/5G software suite},'' \url{https://github.com/srsran/srsRAN}, 2020.

\bibitem{imp4gt}
D.~Rupprecht, K.~Kohls, T.~Holz, and C.~P\"{o}pper, ``{IMP4GT: IMPersonation Attacks in 4G NeTworks},'' in \emph{ISOC Network and Distributed System Security Symposium (NDSS)}.\hskip 1em plus 0.5em minus 0.4em\relax ISOC, Feb. 2020.

\bibitem{lteinspector}
S.~R. Hussain, O.~Chowdhury, S.~Mehnaz, and E.~Bertino, ``{LTEInspector}: {A} {Systematic} {Approach} for {Adversarial} {Testing} of {4G} {LTE},'' in \emph{Proceedings 2018 {Network} and {Distributed} {System} {Security} {Symposium}}.\hskip 1em plus 0.5em minus 0.4em\relax San Diego, CA: Internet Society, 2018.


\bibitem{michaelides2025assessing}
S.~Michaelides, J.~Mucke, and M.~Henze, ``Assessing the Latency of Network Layer Security in 5G Networks,'' in \emph{Proceedings of the 18th ACM Conference on Security and Privacy in Wireless and Mobile Networks (WiSec~'25)}, New York, NY, USA: Association for Computing Machinery, 2025, pp.~262--267, doi: 10.1145/3734477.3734722

\bibitem{ludant2025wiselowlayer}
N.~Ludant, M.~Vomvas, and G.~Noubir,
``Low-Layer Attacks Against 4G/5G Networks,''
in \emph{Proceedings of the 18th ACM Conference on Security and Privacy
in Wireless and Mobile Networks (WiSec~'25)}, ACM, 2025.


\bibitem{zhang2022airinterface}
X.~Zhang, M.~Liyanage, M.~Ylianttila, and A.~Gurtov,
``Improving 4G/5G air interface security: A survey of existing attacks and defenses,''
\emph{Computer Networks}, vol.~212, p.~109032, 2022.

\bibitem{ahmad2022systematic5g}
M.~T.~Alshurideh \emph{et al.},
``A Systematic Literature Review of Security in 5G-based Systems,''
in \emph{Proc. 2019 International Electrical Engineering Congress (iEECON)}, 
IEEE, 2022, doi: 10.1109/ICCR56254.2022.9996068.



\bibitem{nrscope}
H. Wan, X. Cao, A. Marder, and K. Jamieson,
``NR-Scope: A Practical 5G Standalone Telemetry Tool,''
CoNEXT Companion, 2024.


\bibitem{cao2025mac}
J.~Cao, Y.~Yang, R.~Ma, S.~Li, and H.~Li,
``The Security Overview and Analysis of 3GPP 5G MAC CE,''
\emph{arXiv preprint} arXiv:2506.09502, 2025.


\bibitem{vu2024phyply}
M.~Vu, A.~Kalla, and V.~Kozma,
``Survey on 5G Physical Layer Security Threats and Countermeasures,''
\emph{Sensors}, vol.~24, no.~17, p.~5574, Aug.~2024.

\bibitem{hossein2022jamming}
H.~Rahman, M.~H.~M.~Selim, Y.~E.~Baykas, and M.~J.~Hossain,
``Jamming Attacks and Anti-Jamming Strategies in Wireless Networks,''
\emph{IEEE Commun. Surveys Tuts.}, vol.~24, no.~4, pp.~2841--2887, 2022.

\bibitem{zhang2025pls}
S.~Zhang, J.~Liu, Y.~Wu, Z.~Lin, and D.~W.~K. Ng,
``A systematic survey on physical layer security oriented to reconfigurable intelligent surface empowered 6G,''
\emph{Comput. Secur.}, vol.~148, p.~104100, Jan.~2025.


\bibitem{hussain2019reasoner}
S.~R.~Hussain, M.~Echeverria, I.~Karim, O.~Chowdhury, and E.~Bertino,
``5GReasoner: A Property-Directed Security and Privacy Analysis Framework for 5G Cellular Network Protocol,''
in \emph{Proc. ACM Conf. Computer and Communications Security (CCS)}, 2019.

\bibitem{hussain2025controlchaos}
M.~Akon, M.~T.~Toufikuzzaman, and S.~R.~Hussain,
``From Control to Chaos: A Comprehensive Formal Analysis of 5G’s Access Control,''
in \emph{Proc. IEEE Symp. Security and Privacy (S\&P)}, 2025.


\bibitem{networksignalguru}
Network Signal Guru,
``NSG: Multi-functional Android OS tool for voice/data service QoS troubleshooting and RF optimization,''
QTRUN Technologies, Android app v4.7.10, Mar.~2025.



\end{thebibliography}
%

\end{document}